\documentclass[]{aa}
\usepackage{amsmath,graphicx,amssymb}
\usepackage{adjustbox}
\usepackage{txfonts}
\usepackage{natbib}
\usepackage{hyperref}
\newcommand{\vsini}{\ensuremath{{\upsilon}\sin i}}
\newcommand{\kms}{\,km\,s$^{-1}$}
\newcommand{\Teff}{$T_\mathrm{eff}$}
\newcommand{\logg}{$\log g$}

\begin{document}

\title{Geneva seven-colour synthetic photometry from \textit{Gaia} DR3 spectrophotometry}

\author{E.~Paunzen}
\institute{Department of Theoretical Physics and Astrophysics, Masaryk University,
Kotl\'a\v{r}sk\'a 2, 611\,37 Brno, Czechia \\
\email{epaunzen@physics.muni.cz}
}

\date{}

\abstract
{The well-established calibrations of astrophysical parameters derived from classical 
photometric systems (e.g. those of Johnson, Str{\"o}mgren-Crawford, and Vilnius) are rarely applied at present. Instead, spectral energy distribution fitting techniques and very broad $Gaia$ photometric system tend to be employed. However, these approaches have turned out to be highly inaccurate for upper main sequence stars.}
{In this work, the Geneva seven-colour photometric system was synthesised from 
flux-calibrated low-resolution $Gaia$ spectrophotometry (BP/RP spectra). The available ground-based
observations have rendered it to the most homogeneous one available, offering a number of astrophysical
calibrations and applications.}
{The GaiaXPy Tool was used to synthesise the seven filters. The original Geneva $U$ had to be 
exchanged with the SDSS $u$ because it is found to be significantly bluewards of the cut-off
of the BP/RP spectra. The ground-based observations were used to correct for offsets and
the known \Teff-dependency in the ultraviolet region of the BP/RP spectra.}
{The test using well-known \Teff\ and metallicity calibrations on the synthetic photometry
(with 21\,080 stars) demonstrated the high accuracy of this approach. Further checks for possible
correlations with the projected rotational velocity and binarity produced a negative result. As a direct application,
the metallicity of 630 open clusters (and, thus, the radial Galactic metallicity gradient) were determined.
Finally, we compiled a catalog of 25\,881\,397 objects.}
{This work demonstrates that there are alternatives to the widely used three-band $Gaia$
photometric system. The Geneva seven-colour approach can help calibrate and analyse astrophysical parameters, especially those of upper main sequence stars.}


\maketitle

\titlerunning{Geneva 7-colour synthetic photometry}
\authorrunning{Paunzen}

\section{Introduction} \label{introduction}

For many applications in stellar astronomical research, it is important to have accurate estimates of 
astrophysical parameters. Among these parameters, there are three crucial values that allow us to determine the properties of a star and its location in the Hertzsprung-Russell diagram (HRD): (i) effective temperature, \Teff\, in [K], (ii) surface gravity, \logg\, in dex, derived from $g$ in [cm s$^{-2}$], and (iii)  metallicity (either given as [M/H], [Z], or [Fe/H] in solar units or dex). Based on these initial parameters, further parameters such as the age and mass can be determined.

With the dawn of the $Gaia$ satellite, we have seen an inflation in the availability of different techniques to derive stellar astrophysical parameters. In this work, we focus on three typical, widely used methods. The StarHorse2021 \citep{2019AandA...628A..94A,2022A&A...658A..91A} code uses combined parallaxes 
and photometry from the $Gaia$ DR3 together with the photometric catalogues
of Pan-STARRS 1 \citep{2013ApJS..205...20M}, 2MASS \citep{2006AJ....131.1163S}, AllWISE \citep{2013yCat.2328....0C}, 
and SkyMapper DR2 \citet[without the $u$ filter; ][]{2019PASA...36...33O} to derive
Bayesian stellar parameters, distances, and extinctions. In particular, the StarHorse code \citep{2018MNRAS.476.2556Q} is a Bayesian parameter estimation code that compares many observed quantities to stellar evolutionary models. The $Gaia$ DR3 Astrophysical Parameter Inference System \citep[Apsis;][]{2023AandA...674A..28F} is the pipeline software developed by the $Gaia$ consortium. It is used to analyse astrometry, photometry, BP/RP, and RVS spectra for objects across the HRD. 
The LASSO-MLPNet \citet{2022MNRAS.514.4588L} uses a neural network-based scheme to deliver atmospheric parameters from low-resolution stellar spectra of LAMOST. This approach employs a polynomial-fitting method to estimate and remove the pseudo-continuum for each spectrum. Secondly, it has been used to detect parameter-sensitive features using the least absolute shrinkage and selection operator (LASSO). 

However, \citet{2025Astro...4...15K} showed that for upper main sequence stars, the estimations 
of the parameters in the different catalogues are 
off by up to almost $\Delta T_\mathrm{eff} = 10^4$K in terms of the temperature and 2 dex 
in surface gravity. Therefore, the current situation is not satisfactory with respect to precision.

We recall that empirical calibrations of the stellar astrophysical parameters
have existed since the introduction of photometric systems 
\citep[][ and references therein]{1958LowOB...4...37J,1966ARA&A...4..433S,1974ASSL...41.....G}.
Although they are no longer commonly applied, their advantages are manifold. They are well 
tested and based on intermediate- or narrow-band photometric systems, such as the
Str\"omgren-Craford \citep{1993A&A...268..653N}, Geneva \citep{1999NewAR..43..343C} and 
Vilnius \citep{1999ASPC..167..324S} systems. They also include reddening-free indices, which
makes the determination even more robust.

To access these calibrations, synthetic photometry based on the 
$Gaia$ DR3 provides flux-calibrated low-resolution spectrophotometry (BP/RP spectra) can be used.
They provide data
for about 200 million sources in the wavelength range from
330 to 1050\,nm \citep{2023A&A...674A..33G}.
From these spectra, synthetic photometry can be derived for any passband. They have already been used to prepare a wide variety of photometric systems, but caution needs to be taken when checking the 
absolute calibration \citep{2026MNRAS.548ag544P}.

In this paper, a catalogue of synthetic seven-colour Geneva photometry \citep{1980VA.....24..141G} 
for stars hotter than K0 is presented. 
All types of `normal' stars were investigated within the Geneva photometric system, from 
the lower main sequence \citep{1977PASP...89..706E} to the most massive ones 
\citep{1976A&A....48...87G,1993A&A...269..457C}, including giants 
\citep{1989Ap.....31..735K}. Furthermore, binaries \citep{1998A&AS..132..367B}, Be stars \citep{1999A&A...346..134B},
chemically peculiar stars of the upper main sequence \citep{1980A&A....92..289H,1982A&A...114...23H}, 
Population II objects \citep{1986spcp.proc...53H}, 
shell stars \citep{1987A&A...177..193H,2000A&A...354..157H},
variables of all 
types \citep{1981A&A....97..274W,1992A&AS...95..471L,1994A&AS..108....9B,2000A&A...361..201E}, and
weak-lined F-type stars \citep{1991A&A...252..260H} were successfully analysed.

The synthetic data were calibrated using the available ground-based Geneva photometry published by
\citet{2022A&A...661A..89P}. The quality of the final standard magnitudes was checked by comparing
\Teff\, and [Fe/H] values from the observed and synthetic data sets. Further tests for the possible
effects of binarity and the projected rotational velocity were negative.
Finally, we estimated open cluster metallicities from members.

\begin{figure}
\begin{center}
\includegraphics[width=0.48\textwidth]{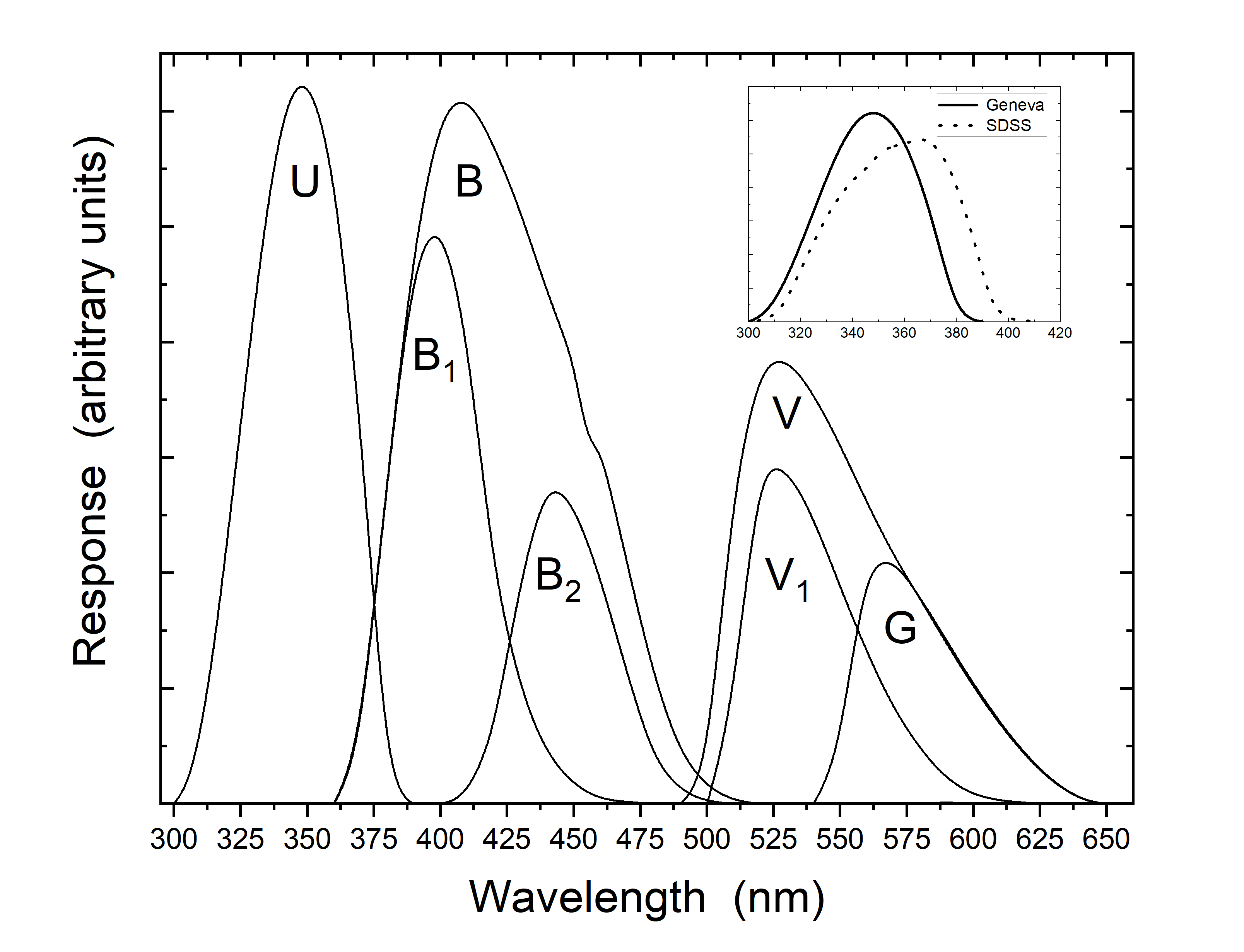}
\caption{Filter curves of the seven filters taken from \citet{1988A&A...206..357R}. 
The inserted figure shows the SDSS $u$ used instead of the original Geneva $U$ one
(see Sect. \ref{sample_calibration}.)}
\label{passbands}
\end{center}
\end{figure}

\section{Characteristics of the Geneva seven-colour photometric system} \label{the_system}

The system was defined in the late 1950s by Marcel Golay and first employed at the Sphinx Observatory 
of the Jungfraujoch Scientific Station
in 1960. It was subsequently applied at the Observatoire de Haute Provence,
the Gornergrat, Calar Alto, and La Silla observatories, and at
the IAC Observatory (Canary Islands). Detailed reviews
can be found in \citet{1972VA.....14...13G}, \citet{1980VA.....24..141G}, 
\citet{1988A&A...206..357R}, and \citet{1999NewAR..43..343C}.

The passbands (Fig. \ref{passbands}) were selected with the aim of mimicking the 
fundamental characteristics of the then-recently developed and notably effective 
Johnson $UBV$ system. Although the $V$ filter closely resembles the Johnson one, 
the $U$ and $B$ passbands have been slightly modified, resulting in a reduced overlap 
at the Balmer jump within the hydrogen spectrum. 
To ensure that the stellar classification attributes of spectrophotometry (which include the 
hydrogen absorption lines and gradients defined over the Paschen continuum) were accurately 
represented, four additional intermediate filters (i.e. $B_\mathrm{1}$, $B_\mathrm{2}$, 
$V_\mathrm{1}$, and $G$) were incorporated. Collectively, these filters cover the wavelength 
range from 300 to 650 nm. Several indices have been defined over the decades and we describe them here. For the specific definitions, we refer to Sect. \ref{catalogue}).

The classical colour indices $U-B_\mathrm{2}$, $B_\mathrm{2}-V_\mathrm{1}$ 
\citep[both insensitive to metallicity,][]{1980VA.....24..141G}, 
and $V_\mathrm{1}-G$ \citep[useful to detect chemically peculiar stars,][]{1982A&A...114...23H}, 
are well correlated to the \Teff\ for different spectral types, but remain sensitive to the 
interstellar reddening.

The indices $g$ and $m_\mathrm{2}$ are similar to the $m_\mathrm{1}$ index of the Str{\"o}mgren-Crawford
system \citep{1966ARA&A...4..433S}. They are sensitive to the metallicity, measuring the 
effect of line blocking in the region redder than 370\,nm. The difference between these two
indices is the definition of how they were made insensitive to interstellar reddening.
Different from $m_\mathrm{1}$, the lines of constant values (isolines) are roughly parallel to 
the \Teff\  axis for \Teff\,$>11\,000$\,K and
are roughly parallel to the \logg\  axis for \Teff\,$<7\,000$\,K.
    
Next we have $d$ and $\Delta$.  These indices are quite comparable to the $c_\mathrm{1}$ index of the 
Str{\"o}mgren-Crawford
system \citep{1966ARA&A...4..433S} and the $Q$ parameter of the 
Johnson system \citep{1975PASP...87..805G}. The $d$ index primarily assesses the Balmer 
discontinuity and is not affected by interstellar extinction (similar to $\Delta$, $g$, and $m_\mathrm{2}$). 
The lines of constant $d$ closely resemble the lines of constant $(U-B)$. 
The $\Delta$ index is influenced by both chemical composition and gravity in cool-type stars.
        
Finally, we have $X$, $Y$, and $Z$. These indices, which are free from reddening effects, were 
developed for stars with temperatures exceeding 8500\,K by rotating the 3D space defined by 
the $d$, $\Delta$, and $g$ indices. Consequently, $X$ shows an optimal correlation with
\Teff\ for B-type stars, while $Y$ correlates with \logg\ within the same spectral type range 
\citep{1979A&A....78..305C}. Nevertheless, it was soon recognised that these indices could 
also differentiate between cooler-type dwarfs and giants \citep{1978prph.book.....G}, as
well as between typical stars and chemically peculiar stars on the upper main sequence, particularly 
through the $Z$ index \citep{1980A&A....88..135C}, for instance. The latter paper also
investigated the relation between surface magnetic field intensities and $Z$ for
B-type stars using the 
520\,nm flux depressions typical for magnetic chemically peculiar stars 
\citep{2003MNRAS.341..849K,2005A&A...441..631P}. They found a clear correlation for mean surface 
densities between 500 and 5000\,G (above that value, $Z$ gets saturated). This result was
extended to cool chemically peculiar stars by \citet{2006A&A...448.1153K}.

\begin{table}
\caption{Statistical
characteristics of the histograms of the differences as shown in Fig. \ref{histograms_differences}.}
\label{table_histograms}
\tiny
\begin{tabular}{ccccccc}
\hline
 &  $\Delta U$ & $\Delta V$ & $\Delta B_{\mathrm 1}$ & $\Delta B_{\mathrm 2}$ & $\Delta V_{\mathrm 1}$ &
 $\Delta G$ \\
\hline
Mean    &       +0.000  & +0.001        & $-$0.001      & +0.001        & +0.000 & +0.002\\
Median  &       +0.000  & +0.000        & +0.000        & +0.000        & +0.000 & 0.000 \\
$\sigma$        &       +0.072  & +0.023        & +0.013        & +0.015        & +0.021 & +0.027\\
\hline
\end{tabular}
\end{table}

\section{The sample selection and calibration} \label{sample_calibration}

The Geneva filters from  
The Spanish Virtual Observatory SVO\footnote{\url{https://svo2.cab.inta-csic.es/theory/fps/}}
were used within the GaiaXPy Tool\footnote{\url{https://gaia-dpci.github.io/GaiaXPy-website/}}.
The Geneva $U$ could not be calculated because it is beyond (blueward) the cut-off of the
BP/RP spectra. Therefore, the SDSS $u$ (Fig. \ref{passbands}), which was already included 
with standard magnitudes in the paper
by \citet{2023A&A...674A..33G}, was taken instead.

Several Geneva indices split into main sequence stars and giants at a spectral type of K0 
\citep{1994ASPC...60..157H}. It was not possible to resolve this ambiguity based on the 
currently available BP/RP spectra. The reason is most probably the absolute flux calibration
in this spectral region. Therefore, a cut at $Y$\,$>$\,$-$0.6 was made \citep[see Fig. 5 in][]{2022A&A...661A..89P}.

As the calibration sample, all stars with photoelectric 
ground-based observations from the most recent catalogue by
\citet{2022A&A...661A..89P} were taken and cross-matched with the synthetic data set. No pre-selection was made to exclude, for instance, binaries, giants, 
low-metallicity (i.e. Population II) stars, known peculiar objects, or highly reddened stars. 
This guarantees no further biases and the universality of the derived
calibration. 
Tracing the widely used threshold value of 1.25 for the RUWE parameter 
\citep{2024A&A...688A...1C} from the $Gaia$ DR3 can help us locate binaries, for example; however, a non-standard value can be explained by (astrometric) binarity, extended stellar
sources (due to circumstellar or protoplanetary disks), crowding in the Galactic disk,
contamination from neighbouring sources, astrophysical variability (due to 
large starspots or flares), and/or data processing or calibration artifacts.  
The parameter `duplicate\_source' indicates close binarity, but 
also instrumental and identification problems. 
In total, 6188 objects have a duplicate\_source value of 1, 
and, additionally, 3740 stars have a RUWE parameter larger than 1.25, respectively. Therefore, 
the sample accounts for all possible instrumental and non-standard effects.

To avoid saturation effects on the BP/RP spectra \citep[see Fig. 46 in][]{2023A&A...674A...3M}, the
sample was limited to $Gaia$ DR3 $G$\,$>$\,6\,mag, respectively. The faintest objects are of the 14th magnitude,
which is due to the brightness limit that could have been reached with the ground-based observations.
In total, 21\,080 stars were used to calibrate the synthetic photometry.

Here, we ecall that since the first catalogue of this photometric system,
all magnitudes were listed with respect to the $B$ filter \citep{1971A&AS....4...43R,1989A&AS...78..469R}. 
Therefore, as the first step, the magnitudes were normalised to the $B$ filter, and 
the differences of the synthetic and observed magnitudes in six filters were derived. As mentioned above,
it is well-known that the UV region of the BP/RP spectra shows some offsets in correlation with the
\Teff\ especially for upper main sequence stars. Therefore, the differences were checked using 
the Johnson $Q$ parameter introduced by \citet{1953ApJ...117..313J}. It is a classical reddening-free
parameter that traces the \Teff\ for a wide range of spectral types 
\citep{1975PASP...87..425G,1975PASP...87..805G}. It is defined as 
\begin{equation}
Q = (U-B) - \frac{E(U-B)}{E(B-V)}(B-V) = (U-B) - 0.72(B-V),
\end{equation}
using the commonly adopted reddening ratio from \citet{1958LowOB...4...37J}. The Johnson $UBV$
photometry for this work 
was taken from the synthetic $Gaia$ photometry as well. The validity of the Johnson $U$
magnitudes was proven using a huge compilation of 5491 open clusters \citep{2026arXiv261107609P}.
A quadratic relation was identified and corrected as
\begin{equation}
U_\mathrm{std} - U_\mathrm{synth} = +0.065(1) + (0.175(6)*Q) - (0.209(8)*Q^2).
\end{equation}

\begin{figure}
\begin{center}
\includegraphics[width=0.48\textwidth]{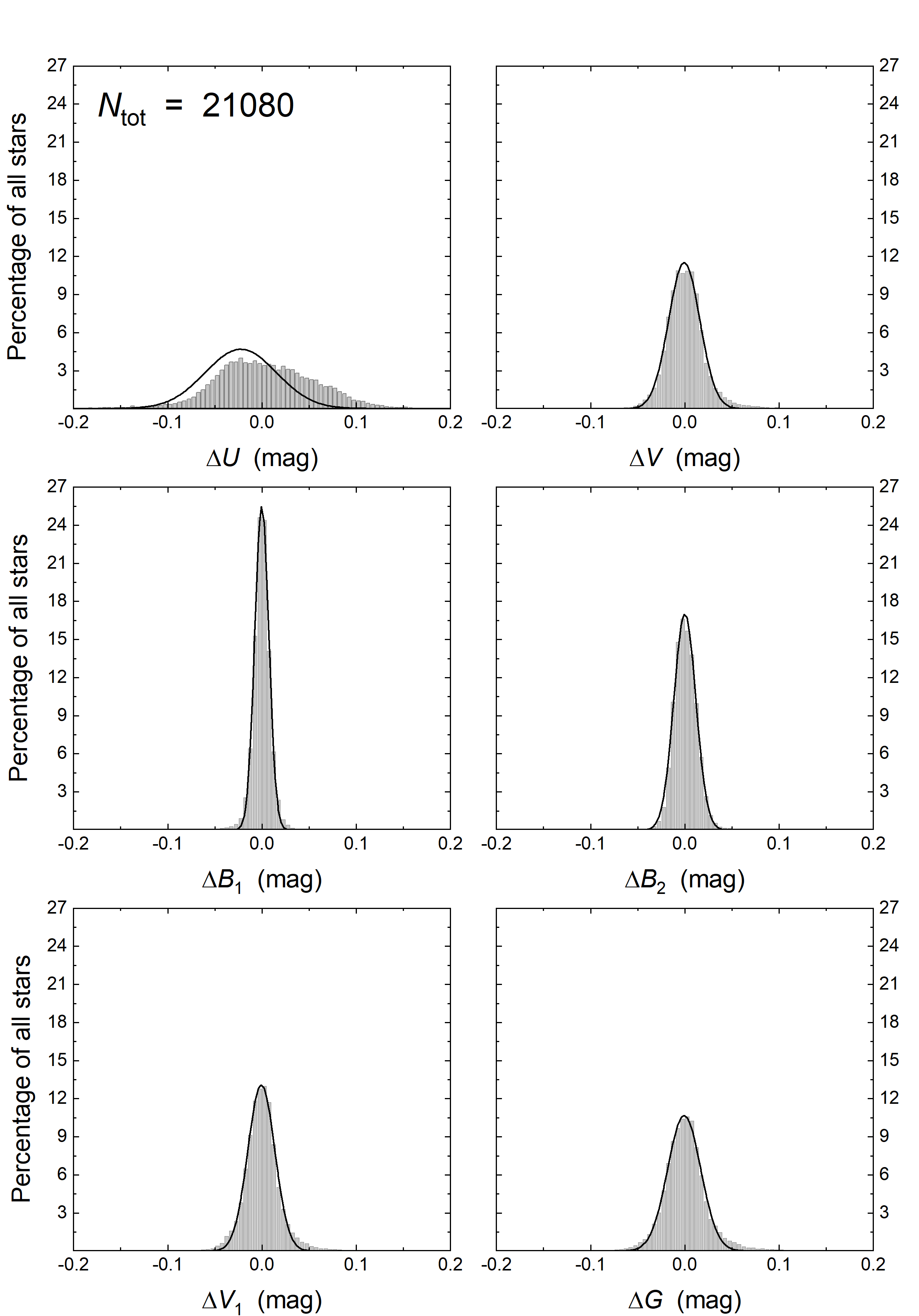}
\caption{Histograms of the differences between the observed and synthesised samples as 
discussed and the final calibrated values derived in Sect. \ref{sample_calibration}.
The statistical properties are listed in Table \ref{table_histograms} and the
Gaussian distributions overplotted. The scaling for
all plots is identical.}
\label{histograms_differences}
\end{center}
\end{figure}

None of the other filters shows any correlation
with $Q$, respectively. In Fig. \ref{histograms_differences}, the histograms of the differences 
between the observed and synthesised samples are shown. The statistical properties of the 
distributions are listed in Table \ref{table_histograms}. All panels share the same scaling. It is obvious that
the distribution of $U$ is different compared to the other ones. It is much broader and not symmetric.
All indices, which also include $U$ were checked for systematic trends with \Teff, metallicity, and 
\logg, but none were detected. 

\begin{figure}
\begin{center}
\includegraphics[width=0.48\textwidth]{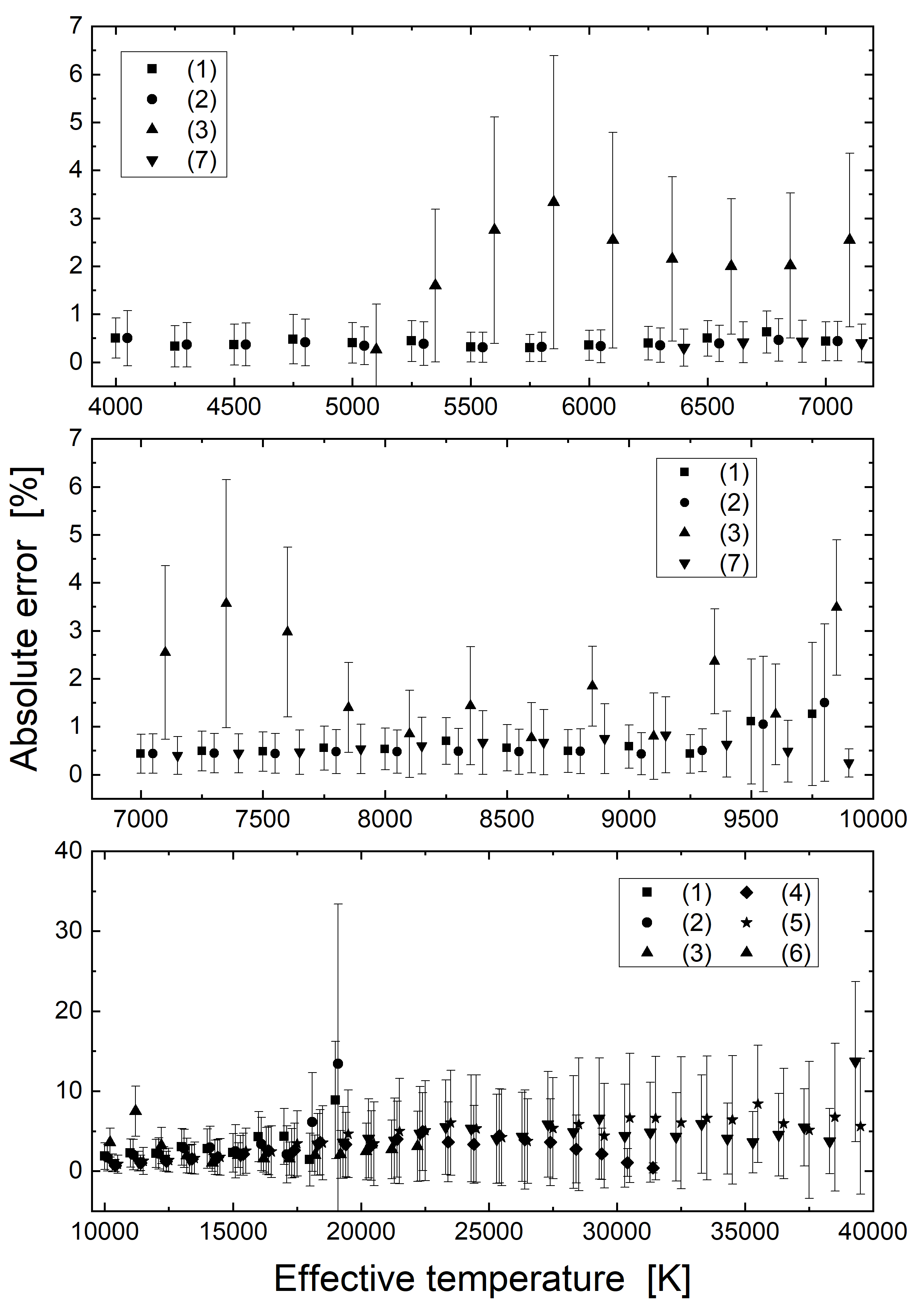}
\caption{Results of the different \Teff\ calibrations for the observed and synthetic 
photometric magnitudes. The calibrations are (1) \citet{1996BaltA...5..303H} for $B_\mathrm{2}-G$;
(2) \citet{1996BaltA...5..303H} for $B_\mathrm{2}-V_\mathrm{1}$; (3) \citet{1997A&AS..122...51K};
(4) \citet{1999NewAR..43..343C} for \logg\,=\,3.5; (5) \citet{1999NewAR..43..343C} for \logg\,=\,5.0;
(6) \citet{2005A&A...444..941P}; (7) \citet{2006A&A...458..293P}.}
\label{figure_binning}
\end{center}
\end{figure}

\section{The calibration of the synthetic photometry} \label{calibration}

To test the quality of the synthetic photometry over the complete spectral range, the following 
\Teff\ calibrations were used: 

\begin{itemize}
    \item \citet{1996BaltA...5..303H}: $B_\mathrm{2}-G$ and $B_\mathrm{2}-V_\mathrm{1}$ for the complete spectral range;
    \item \citet{1997A&AS..122...51K}: A combination of different indices ($B_\mathrm{2}-V_\mathrm{1}$, $d$,
    $X$, and $Y$) taking into account various \logg\ and metallicity values for B- to G-type stars;
    \item \citet{1999NewAR..43..343C}: $X$ for different \logg\  grids of late O- to late B-type stars;
    \item \citet{2005A&A...444..941P}: $X$ for B-type stars;
    \item \citet{2006A&A...458..293P}: $B_\mathrm{2}-V_\mathrm{1}$ for A- to mid F-type stars.
\end{itemize}

\begin{figure}
\begin{center}
\includegraphics[width=0.48\textwidth]{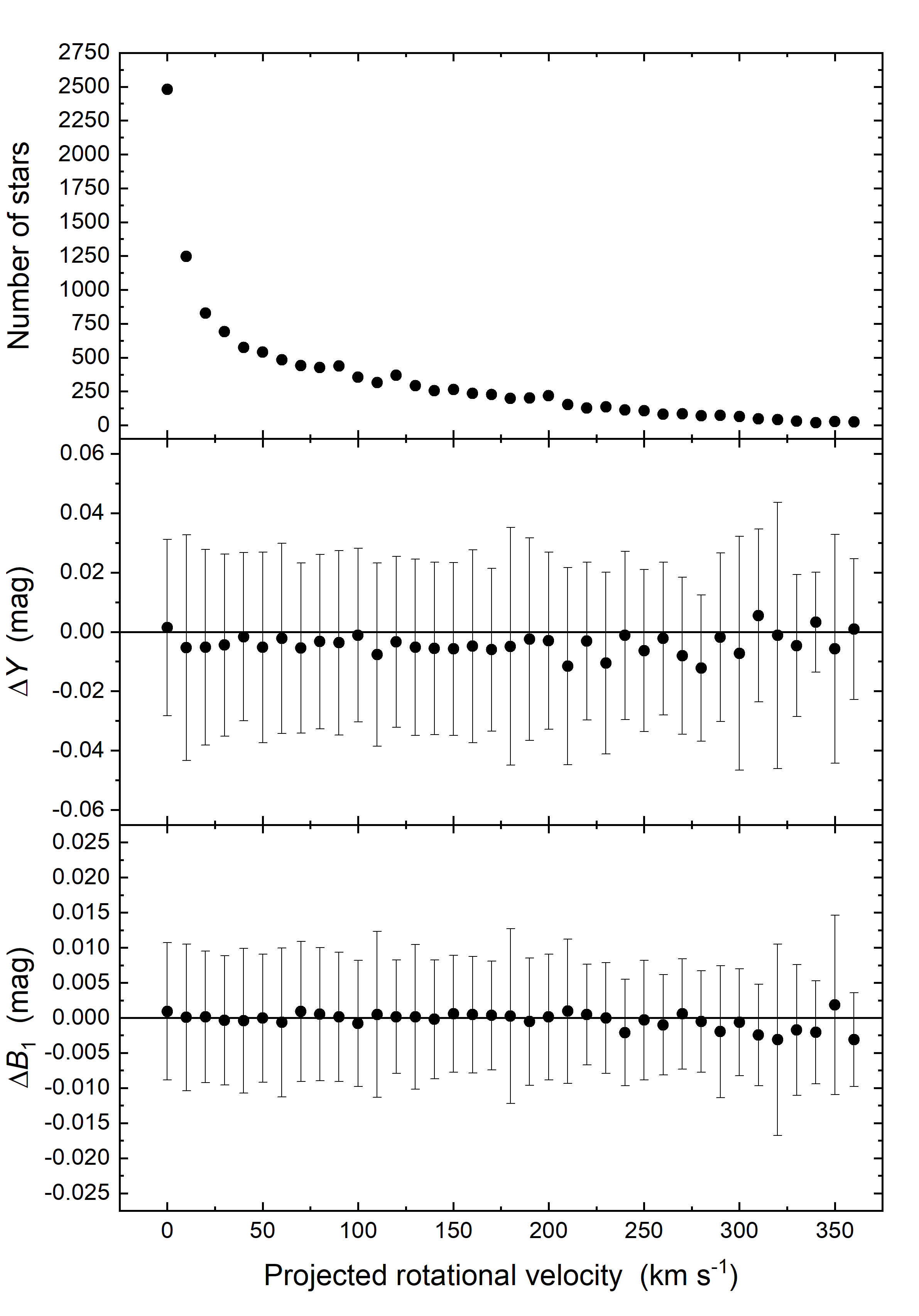}
\caption{Differences between the synthetic and observed magnitudes for the 12\,334 stars with 
known projected rotational velocities (\vsini) from the literature. Upper panel shows the
number of stars per bin. No significant correlations were found for all investigated magnitudes
and indices.}
\label{figure_vsini}
\end{center}
\end{figure}

When applying the calibrations except those using $X$, an estimate of the individual interstellar
reddening is needed. Interstellar reddening, caused by wavelength-dependent extinction by dust 
grains in the interstellar medium, plays a fundamental role in shaping the observed 
properties of Galactic stellar populations
\citep{2003ARA&A..41..241D}. An accurate reddening determination is essential for deriving reliable 
stellar parameters. For the objects of the calibration sample, the reddening values by
\citet{2022A&A...658A..91A} were used. For the transformation of the absorption in $V$ to the corresponding
Geneva colours, the following relations \citep{1976PASP...88..917C} were applied
$A_{\rm V}$\,=\,3.57$E(B_\mathrm{2}-V_\mathrm{1})$\,=\,2.64$E(B_\mathrm{2}-G)$. 

For each star, the measured and synthetic colours were used to derive \Teff\ values according to
the validity of the corresponding calibration. For example, a G-type star was not calibrated using
the relations derived for B-type stars by \citet{2005A&A...444..941P}. Therefore, the number
of objects for the individual calibrations varies. For the calibration by \citet{1999NewAR..43..343C}, the two grids with \logg\,=\,3.5 and
\logg\,=\,5.0 were applied.
Finally, the results were binned according to
the \Teff\ values and absolute errors and their standard deviations were derived. The results are
listed in Table \ref{results_binning_Teff} and shown in Fig. \ref{figure_binning}. It is important to note the
current situation of available \Teff\ calibrations. With the availability of the $Gaia$ data, most
new efforts to calibrate stellar astrophysical parameters concentrated on lower main sequence stars, namely, stars with convective atmospheres and spectral types cooler than F2\,V 
\citep{2021MNRAS.507.2684C,2021A&A...653A..90M}. This is for several reasons: most importantly the 
decreasing signal-to-noise (S/N) of the BP/RP spectrophotometry in the blue wavelength region  
\citep{2023A&A...674A...3M}. The  $Gaia$ $BP-RP$ colour index that is commonly used today is close to
insensitive for hot and massive stars. Another important fact is that the peak of the 
spectral-energy distribution for an A0 star is already in the ultraviolet region, which is not observed 
by $Gaia$.

From Fig. \ref{figure_binning} it can be concluded that the calibrations of \citet{1996BaltA...5..303H}
result in differences below 1\% for spectral types up to A1\,V (or about 9300\,K). Then, the differences
slightly increase up to 3\% for stars of B3\,V (17\,000\,K). The differences for A-type stars
calibrated with the formulae for A- to mid F-type stars are below 1\%, respectively.
The calibration by \citet{1997A&AS..122...51K}
gives in the spectral range from K2\,V to B3\,V variations between 1\% and 4\%. This is caused by the applied grids and their steep gradients, which are very sensitive to slight differences in the individual colours. 
The calibrations by \citet{1999NewAR..43..343C} and \citet{2005A&A...444..941P} for B-type stars
show an increase of the differences from 1\% to about 6\%. The latter is caused by the increasing 
insensitivity of $X$ with \Teff\ for the most massive stars. Nevertheless, the calibrations of the
synthetic photometric colours can be applied for stars as hot as 39\,000\,K with a very high
accuracy.

As an additional test, the metallicity calibration for early A- to K-type 
(9000 to 5000\,K) stars published by
\citet{2017MNRAS.469.3042N} was applied. He used the $m_\mathrm{2}$ and $B_\mathrm{2}-V_\mathrm{1}$
indices to establish a standard line. The deviation of the standard line $\Delta m_\mathrm{2}$
was then correlated with published [Fe/H] values. In addition, the $t$/$U$ plane was used, where
$t = (B_\mathrm{2}-G) - 0.39(B_\mathrm{1}-B_\mathrm{2})$ is a good temperature indicator for
cool type stars \citep{2003A&A...398..705M}. The calibration is valid for 
$-$2.5\,$<$\,[Fe/H]\,$<$\,+0.5\,dex, respectively. As a test, the bin of 5750\,K with 1238
stars was taken and divided into individual [Fe/H] bins of 0.25\,dex. No correlation 
of the differences with the absolute [Fe/H] was found. Therefore, for the 
test, no discrimination of the 
absolute [Fe/H] values per bin was taken into account. For the quality assessment, the differences
of the calibrated [Fe/H] values using the observed and synthetic photometry were used. Table
\ref{results_binning_FeH} lists the results for all bins. The mean differences are between
0.04 and 0.09\,dex for all temperature bins and the complete [Fe/H] range. Again, proof that
the synthetic colours are very accurate and that the Geneva photometry is well-suited to derive
metallicities.

\section{The effect of binarity and rotation} \label{binarity_rotation}

Another important aspect is the influence of binarity and rotation on photometric systems 
and indices. A fact which is nowadays widely neglected. Here, we have to distinguish between 
the effect on (1) the calibration of synthetic magnitudes and (2) the effect of colours 
with respect to standard magnitudes of non-rotating stars. For the purposes of this work, the first aspect was investigated.

We recall that \citet{1970A&A.....7..120M} 
and \citet{1989A&A...214..153H} investigated
the effects of rotation on colours and indices of the Geneva photometric system (effect (2) from above).  
They found that the ($X$, $Y$) and ($d$, $\Delta$) planes for B- and A-type stars are not significantly
affected by rotation. However, there is a specific effect that acts on the classical colour indices: 
$\Delta(B_\mathrm{2}-V_\mathrm{1})$\,=\,k(\vsini)$^2$. Such an effect is also widely recognised in the literature
for the Str{\"o}mgren-Crawford $(b-y)$ colour, for example: \citep{1987ApJS...65..581G}.
If an effect of this sort also exists for other \Teff\ sensitive indices such as $B_\mathrm{2}-G$, it will be the subject of a subsequent paper. 

To investigate the possible effect of \vsini, the $Gaia$ vbroad and our own compilation 
of \vsini\ values from the literature were used. The $Gaia$ vbroad is an estimation 
of \vsini\ from the $Gaia$ RVS spectrograph \citep{2023A&A...674A...8F}. 
It performs well for velocities between 75 and 200 \kms, but underestimates the \vsini\ 
of slow rotators (\vsini\,$<$\,50\kms) and overestimates it for rapid rotators 
\citep[\vsini\,$>$\,200\kms;][]{2024MNRAS.534.3022M}. Whenever possible, mean \vsini\ values
were calculated. In total, estimates for 12\,334 stars up to 370\,\kms\ were found. 
The individual values were binned into 10\,\kms\ bins regardless of the stellar 
colour (\Teff, or spectral type). 

Figure \ref{figure_vsini} shows the differences between the synthetic and observed magnitudes
for the $B_\mathrm{1}$ magnitude (lower panel) and the $Y$ index (middle panel). They were solely chosen as examples. None of the derived synthetic magnitudes reveal any effect with \vsini.
This conclusion is based on a large and statistically sound sample of stars over the whole
\vsini\ and spectral type range.

Unresolved binarity affects any photometric magnitude caused by the combination of the flux of two 
(or more) stars scaled by their luminosity ratio. Again, we find the effect for the calibration
itself and the change of the individual values with respect to single stars. The latter was
extensively studied by \citet{1999NewAR..43..343C}. He calculated binarity loops for
different indices (Figs. 14 to 16 therein). 

As stated in Sect. \ref{sample_calibration}, 9319 objects of our calibration sample can be
classified as binaries according to the $Gaia$ DR3 (Duplicity flag and the RUWE value). 
To check the possible influence of binarity, the calibration procedure described in 
Sect. \ref{calibration} was applied to the sample of apparent single stars and binaries only.
The coefficients and offsets together with their errors were then compared with each other and
the ones from the combined (final) overall sample. Within the standard deviations, no statistically
significant differences were found. 

\begin{figure}
\begin{center}
\includegraphics[width=0.48\textwidth]{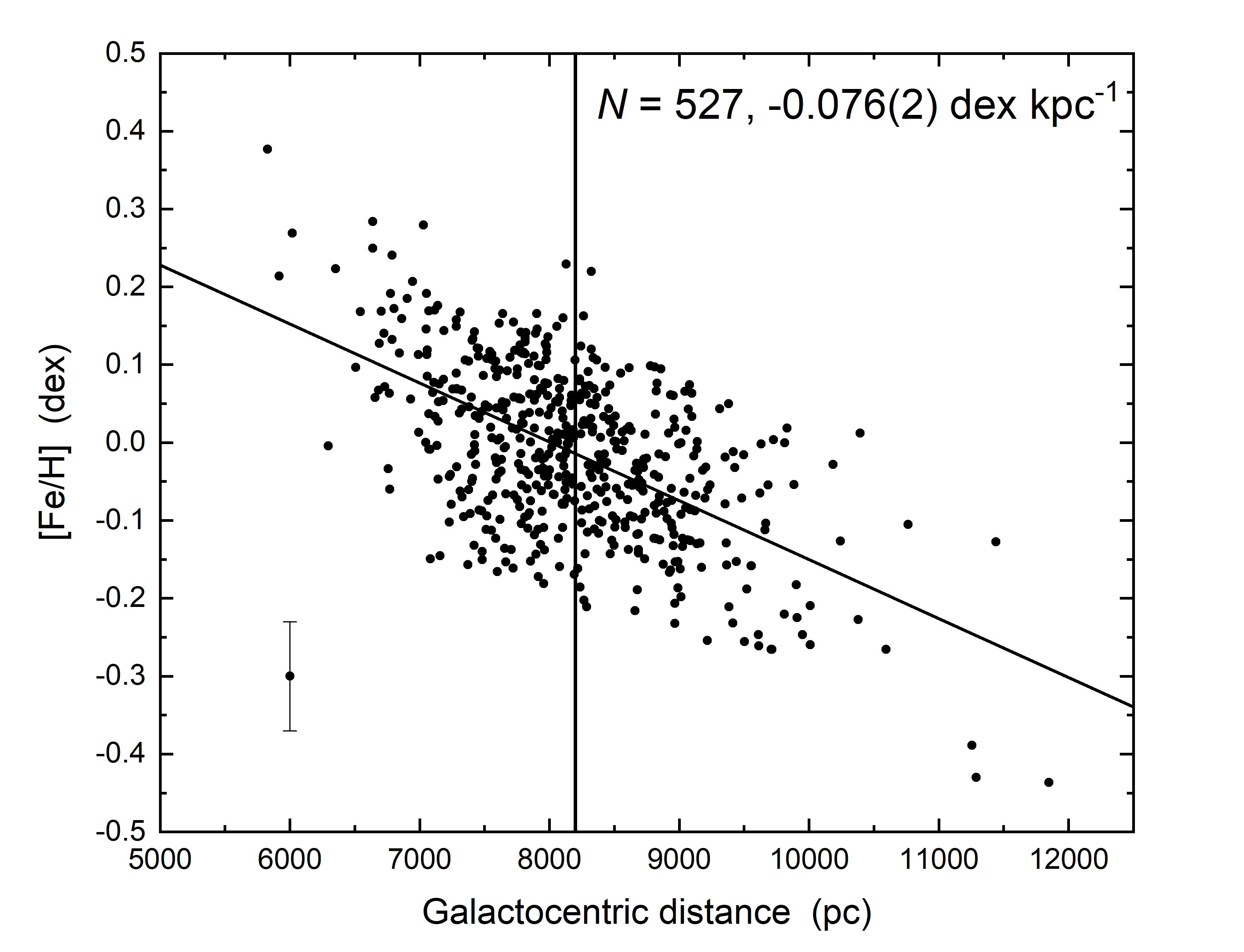}
\caption{Calibrated [Fe/H] values of open clusters using the calibration by
\citet{2017MNRAS.469.3042N} for individual members versus the galactocentric distance.
The vertical line is the location of the Sun at 8200\,pc. The point at the lower left
side shows the typical error of the individual values. The gradient was determined
as $-$0.076(2)\,dex\,kpc$^{-1}$.}
\label{figure_rgc}
\end{center}
\end{figure}

\section{The catalogue and its usage} \label{catalogue}

The number of stars included in the catalogue is much lower than the overall number of available
BP/RP spectra (i.e. about 200 million). This results from the limit of the $U$ magnitude, which
is much brighter than in the case of the other filters. The errors of colours were included as they are calculated by GaiaXPy Tool for the individual fluxes. They were converted as $\Delta mag = 1.0857 \Delta flux$ and for all indices a full error propagation was done. We notice that the magnitudes were normalised to the $B$ 
filter as in the original photoelectric system. In particular, the table is organised as follows: 

\begin{itemize}
\item Column 1: $Gaia$ identification;
\item Column 2: Right ascension (J2016.0);
\item Column 3: Declination (J2016.0);
\item Column 4: Galactic longitude;
\item Column 5: Galactic latitude;
\item Column 6: $Gaia$ DR3 $G$ magnitude;
\item Column 7: Johnson $Q$;
\item Column 8: Geneva $G-B$;
\item Column 9: $\sigma$ of Geneva $G-B$;
\item Column 10: Geneva $U-B_\mathrm{1}$;
\item Column 11: $\sigma$ of Geneva $U-B_\mathrm{1}$;
\item Column 12: Geneva $U-B_\mathrm{2}$;
\item Column 13: $\sigma$ of Geneva $U-B_\mathrm{2}$;
\item Column 14: Geneva $B_\mathrm{1}-B_\mathrm{2}$;
\item Column 15: $\sigma$ of Geneva $B_\mathrm{1}-B_\mathrm{2}$;
\item Column 16: Geneva $V_\mathrm{1}-G$'
\item Column 17: $\sigma$ of Geneva $V_\mathrm{1}-G$;
\item Column 18: Geneva $B_\mathrm{2}-G$;
\item Column 19: $\sigma$ of Geneva $B_\mathrm{2}-G$;
\item Column 20: Geneva $B_\mathrm{2}-V_\mathrm{1}$;
\item Column 21: $\sigma$ of Geneva $B_\mathrm{2}-V_\mathrm{1}$;
\item Column 22: Geneva $d = (U-B_\mathrm{1}) - 1.430(B_\mathrm{1}-B_\mathrm{2})$;
\item Column 23: $\sigma$ of Geneva $d$;
\item Column 24: Geneva $\Delta = (U-B_\mathrm{2}) - 0.832(B_\mathrm{2}-G)$;
\item Column 25: $\sigma$ of Geneva $\Delta$;
\item Column 26: Geneva $g = (B_\mathrm{1}-B_\mathrm{2}) - 1.357(V_\mathrm{1}-G)$;
\item Column 27: $\sigma$ of Geneva $g$;
\item Column 28: Geneva $m_\mathrm{2} = (B_\mathrm{1}-B_\mathrm{2}) - 0.457(B_\mathrm{2}-V_\mathrm{1})$;
\item Column 29: $\sigma$ of Geneva $m_\mathrm{2}$.
\item Column 30: Geneva $X = +0.3788 + 1.3764U - 1.2162B_\mathrm{1} - 0.8498B_\mathrm{2} - 0.1554V_\mathrm{1} + 0.8450G$;
\item Column 31: $\sigma$ of Geneva $X$;
\item Column 32: Geneva $Y = -0.8288 + 0.3235U - 2.3228B_\mathrm{1} + 2.3363B_\mathrm{2} + 0.7495V_\mathrm{1} - 1.0865G$.
\item Column 33: $\sigma$ of Geneva $Y$;
\item Column 34: Geneva $Z = -0.4572 + 0.0255U - 0.1740B_\mathrm{1} + 0.4696B_\mathrm{2} - 1.1205V_\mathrm{1} + 0.7994G$;
\item Column 35: $\sigma$ of Geneva $Z$.
\end{itemize}

Due to the large number of columns, no example is provided here. 
The application of the individual magnitudes and indices can be done using the same procedure as that used for the photoelectric catalogue. The results from Sects. \ref{calibration} and 
\ref{binarity_rotation} can be seen as a guideline for the corresponding spectral range.
Caution needs to be taken for indices including the $U$ filter. Statistical errors
for the astrophysical parameters can be derived from Fig. \ref{figure_binning}.
Otherwise, there are no limitations.

\section{The metallicity of open clusters} \label{metal_OCLs}

As one important application of the synthetic Geneva photometry, the metallicities of open
clusters and the local radial Galactic metallicity gradient were determined. 
The latter is a vital diagnostic for understanding a galaxy's star formation, 
gas accretion, and chemical enrichment over time. Several models were developed,
starting either with a pre-enriched gas disk, 
which then evolves faster in the central parts, thereby steepening the gradient over time
\citep{2001ApJ...554.1044C}. One alternative is to begin with a primordial composition disk that 
contains some metallicity at the centre, resulting in a steep initial gradient. Over time, as 
stars form in the outer regions, this gradient gradually becomes flatter \citep{2016A&A...588A..91M}.
It is also well known that radial mixing through heating and migration flattens the observed 
abundance gradients with age \citep{2015A&A...580A.126K}.

In recent years, several studies have presented radial Galactic metallicity gradients on
the basis of open clusters \citep{2022MNRAS.509..421N,2025AJ....169..214Y,2026AJ....171...91O}.
The results are still affected by large uncertainties in determining cluster parameters,
in general. However, most authors agree that a gradient of $-$0.05 to $-$0.1\,dex\,kpc$^{-1}$ exists
up to a Galactic centre distance of 10\,kpc \citep{2022MNRAS.509..421N}. The results beyond this 
distance significantly deviate from each other.

To have a good estimation of the cluster's parameters (i.e. age, reddening, distance, and central coordinates), 
the published values from the following five
references were averaged: \citet{2022MNRAS.509..421N}, \citet{2023A&A...673A.114H},
\citet{2024A&A...689A..18A}, \citet{2024AJ....167...12C}, and \citet{2025AJ....170..288L}.
Unfortunately, the individual deviations of the ages for the same open clusters are
still as large as in the pre-$Gaia$ era \citep{2015A&A...582A..19N}. However, the mean ages
were only used to sort out open clusters with ages younger than 50\,Myr. This was done
because the applied metallicity calibration is valid only for early A- to K-type stars and is not
usable for pre-main sequence objects. As a next step, stars with membership probabilities 
larger than 50\%  listed by either of the following two references were used. 

\textit{\citet{2023A&A...673A.114H}}: this stupdy applied the Hierarchical Density-Based Spatial Clustering 
of Applications with Noise (HDBSCAN) algorithm to recover star clusters. The authors validated the found aggregates using a statistical density test and a Bayesian convolutional 
neural network for colour-magnitude diagram classification. 

\textit{\citet{2023MNRAS.526.4107P}\footnote{\url{https://ucc.ar/}}}: this study uses a fast membership probabilities (FASTMP) approach to
estimate the membership probabilities. It can be run in supervised and unsupervised modes.
It is based on pyUPMASK \citep{2021A&A...650A.109P}. 

Based on the results of the paper by \citet{2026arXiv261107609P},
which presented extinction values and reddening ratios for 5491 Galactic open clusters using
synthetic Johnson $U$ and $Gaia$ $GBPRP$ photometry, only clusters with more
than seven members were considered for further analysis. These authors showed that this 
limit avoids spurious results due to poor number statistics.
The averaged reddening values (without weighting), central coordinates, 
and distances were then used to calibrate
mean [Fe/H] values and calculate the galactocentric distance ($R_\mathrm{GC}$)
for the distance of the Sun to the Galactic centre of 8200\,pc. The 
mean [Fe/H] values were not weighted because the individual errors (and, thus, weights)
were almost all on the same level. Weighting did not improve the overall errors,
nor did the mean values change significantly. The derived metallicity gradient was
unchanged. In total,
the [Fe/H] values of 527 open clusters with more than five members (14\,891 individual 
stars in total) in the valid spectral type range were determined.

Figure \ref{figure_rgc} shows the result. Due to the limitations of the metallicity calibration and the 
apparent magnitude of the synthetic photometry, mostly star clusters with 
6500\,$<$\,$R_\mathrm{GC}$\,$<$\,10\,000\,pc (corresponding to a radius of 2000\,pc 
around the Sun) were observed.
The gradient was determined as $-$0.076(2)\,dex\,kpc$^{-1}$. While it has been reported
otherwise in some of the literature \citep{2022MNRAS.509..421N}, no differences among different age groups could be detected. In this regard, significant efforts have to be made going forward to obtain more accurate and reliable age determinations of star clusters. 

\section{Conclusions} \label{conclusions}

Although we have made a lot of progress in deep all-sky surveys from the ground and space, collecting
photometry and spectroscopy in a wide wavelength range, we still face significant shortcomings in
determining the astrophysical parameters of upper main sequence stars. The broad three-band $Gaia$ 
photometric system, which is now almost exclusively used, is derived from the actually observed BP/RP
spectrophotometry and was modified in consecutive $Gaia$ data releases. Also, the next data
release will include a newly defined so-called intrinsic photometric system. However, for many applications,
this photometric system is probably insufficient. However, there is, in fact, a way out of this dilemma.

One possible solution is to synthesise and calibrate existing photometric systems. The $Gaia$ consortium has already made significant efforts to synthesise and calibrate several requently used photometric systems\footnote{\url{https://gaia.aip.de/metadata/gaiadr3/synthetic_photometry_gspc/}},
but unfortunately, these are not commonly applied in most stellar astrophysical fields. We can only speculate about the reasons for this, however, it goes beyond the scope of this work.

In this paper, the important and highly homogeneous 
Geneva seven-colour photometric system was added to the list of those synthesised
from $Gaia$ spectrophotometry. This photometric system was successfully applied to investigate
a significant amount of very different stellar groups and analyse a range of concerns. Finally, there are many existing calibrations for this photometric system.

The synthetic photometry was calibrated using 21\,080 objects, including
binaries, giants, low-metallicity (Population II), and known peculiar or highly reddened stars.
The final calibrated data were tested using various \Teff\ and metallicity calibrations, yielding
an excellent accuracy and consistency. No influence of the binarity (noting that the sample includes
9319 apparent binaries) or the projected rotational velocity (whereby estimates for 12\,334 objects are
available) on the calibration was detected. The final catalogue includes 25\,881\,397 entries
for spectral types earlier than K0.

As a first application of the new synthetic photometry, the open cluster metallicities were determined
using published members. In total, the [Fe/H] values of 630 open clusters with more than five members
were derived, while the radial Galactic metallicity gradient as $-$0.076(2)\,dex\,kpc$^{-1}$ was determined.
This value is well in line with the literature. 

For the future, we can think of many applications, such as investigating the parameters of star clusters and
individual stars, along with the Galactic reddening law, and extending the analysis to extragalactic systems.
In addition, some new models and calibrations, including different metallicities and rotation rates, could be
generated and tested. 

\section*{Data availability}

The final catalogue includes 25\,881\,397 entries and will be provided in electronic form 
t the CDS via anonymous ftp to cdsarc.u-strasbg.fr (130.79.128.5) or 
via http://cdsweb.u-strasbg.fr/cgi-bin/qcat?J/A+A/
and the General Catalogue of 
Photometric data \citep[GCPD,][]{1997A&AS..124..349M}\footnote{\url{http://gcpd.physics.muni.cz/}}

\begin{acknowledgements}
This research was carried out within the institutional support framework 
for the development of the
research organization of Masaryk University. This paper is dedicated to Michael Maitzen,
who died during its preparation. He was a great mentor and friend.
This work has made use of data from the European Space Agency (ESA) mission 
{\it Gaia} (\url{https://www.cosmos.esa.int/gaia}), processed by the {\it Gaia} Data 
Processing and Analysis Consortium (DPAC, \url{https://www.cosmos.esa.int/web/gaia/dpac/consortium}). 
Funding for the DPAC has been provided by national institutions, in particular the institutions
participating in the {\it Gaia} Multilateral Agreement. This research has made use of the SIMBAD database,
operated at CDS, Strasbourg, France and 
of the WEBDA data base, operated at the Department of Theoretical
Physics and Astrophysics of the Masaryk University.
\end{acknowledgements}

\bibliographystyle{aa}
\bibliography{main}

\begin{appendix} 

\section{Binned results of the \Teff\ and metallicity calibrations}

\begin{table}
\caption{Result of the metallicity calibration published by \citet{2017MNRAS.469.3042N} 
for the observed and synthetic photometric magnitudes. }
\begin{center}
\label{results_binning_FeH}
\begin{tabular}{cccc}
\hline
(1) & (2) & (3) & (4) \\
\hline
5000    &       89      &       0.09    &       0.10    \\
5250    &       178     &       0.08    &       0.06    \\
5500    &       756     &       0.05    &       0.03    \\
5750    &       1238    &       0.05    &       0.03    \\
6000    &       980     &       0.07    &       0.04    \\
6250    &       984     &       0.05    &       0.03    \\
6500    &       739     &       0.06    &       0.03    \\
6750    &       532     &       0.05    &       0.03    \\
7000    &       548     &       0.04    &       0.03    \\
7250    &       513     &       0.06    &       0.04    \\
7500    &       434     &       0.06    &       0.05    \\
7750    &       334     &       0.05    &       0.05    \\
8000    &       257     &       0.04    &       0.04    \\
8250    &       132     &       0.07    &       0.06    \\
8500    &       68      &       0.04    &       0.06    \\
8750    &       35      &       0.05    &       0.06    \\
\hline
\end{tabular}
\tablefoot{The first column is the \Teff\ bin. 
The second column is the number of stars per bin, followed by the mean difference in [dex] and its standard
deviation in [dex].}
\end{center}
\end{table}



\begin{table*}
\caption{Results of the different \Teff\ calibrations for the observed and synthetic 
photometric magnitudes. }
\begin{center}
\begin{adjustbox}{max width=1.2\textwidth,angle=90}
\label{results_binning_Teff}
\tiny
\begin{tabular}{cccccccccccccccccccccc}
\hline
(1) & (2) & (3) & (4) & (5) & (6) & (7) & (8) & (9) & (10) & (11) & (12) & (13) & (14) & (15) & (16) 
& (17) & (18) & (19) & (20) & (21) & (22)  \\ 
\hline
4000    &       23      &       0.5     &       0.4     &       29      &       0.5     &       0.6     &               &               &               &               &               &               &               &               &               &               &               &               &               &               &               \\
4250    &       30      &       0.3     &       0.4     &       27      &       0.4     &       0.5     &               &               &               &               &               &               &               &               &               &               &               &               &               &               &               \\
4500    &       54      &       0.4     &       0.4     &       64      &       0.4     &       0.4     &               &               &               &               &               &               &               &               &               &               &               &               &               &               &               \\
4750    &       96      &       0.5     &       0.5     &       103     &       0.4     &       0.5     &               &               &               &               &               &               &               &               &               &               &               &               &               &               &               \\
5000    &       171     &       0.4     &       0.4     &       177     &       0.3     &       0.4     &       4       &       0.3     &       1.0     &               &               &               &               &               &               &               &               &               &               &               &               \\
5250    &       243     &       0.4     &       0.4     &       261     &       0.4     &       0.5     &       32      &       1.6     &       1.6     &               &               &               &               &               &               &               &               &               &               &               &               \\
5500    &       790     &       0.3     &       0.3     &       921     &       0.3     &       0.3     &       66      &       2.8     &       2.4     &               &               &               &               &               &               &               &               &               &               &               &               \\
5750    &       1373    &       0.3     &       0.3     &       1320    &       0.3     &       0.3     &       110     &       3.3     &       3.1     &               &               &               &               &               &               &               &               &               &               &               &               \\
6000    &       1153    &       0.4     &       0.3     &       1109    &       0.3     &       0.3     &       209     &       2.5     &       2.2     &               &               &               &               &               &               &               &               &               &               &               &               \\
6250    &       1151    &       0.4     &       0.4     &       1145    &       0.4     &       0.4     &       505     &       2.2     &       1.7     &               &               &               &               &               &               &               &               &               &       125     &       0.3     &       0.4     \\
6500    &       972     &       0.5     &       0.4     &       953     &       0.4     &       0.4     &       911     &       2.0     &       1.4     &               &               &               &               &               &               &               &               &               &       326     &       0.4     &       0.4     \\
6750    &       839     &       0.6     &       0.4     &       799     &       0.5     &       0.4     &       866     &       2.0     &       1.5     &               &               &               &               &               &               &               &               &               &       396     &       0.4     &       0.4     \\
7000    &       885     &       0.4     &       0.4     &       876     &       0.4     &       0.4     &       679     &       2.5     &       1.8     &               &               &               &               &               &               &               &               &               &       487     &       0.4     &       0.4     \\
7250    &       902     &       0.5     &       0.4     &       918     &       0.4     &       0.4     &       389     &       3.6     &       2.6     &               &               &               &               &               &               &               &               &               &       554     &       0.4     &       0.4     \\
7500    &       907     &       0.5     &       0.4     &       905     &       0.4     &       0.4     &       271     &       3.0     &       1.8     &               &               &               &               &               &               &               &               &               &       595     &       0.5     &       0.5     \\
7750    &       893     &       0.6     &       0.5     &       911     &       0.5     &       0.5     &       220     &       1.4     &       0.9     &               &               &               &               &               &               &               &               &               &       597     &       0.5     &       0.5     \\
8000    &       928     &       0.5     &       0.4     &       931     &       0.5     &       0.5     &       177     &       0.8     &       0.9     &               &               &               &               &               &               &               &               &               &       682     &       0.6     &       0.6     \\
8250    &       952     &       0.7     &       0.5     &       967     &       0.5     &       0.5     &       204     &       1.4     &       1.2     &               &               &               &               &               &               &               &               &               &       703     &       0.7     &       0.7     \\
8500    &       1097    &       0.6     &       0.5     &       1126    &       0.5     &       0.5     &       175     &       0.8     &       0.7     &               &               &               &               &               &               &               &               &               &       775     &       0.7     &       0.7     \\
8750    &       1322    &       0.5     &       0.4     &       1303    &       0.5     &       0.5     &       180     &       1.8     &       0.8     &               &               &               &               &               &               &               &               &               &       815     &       0.8     &       0.7     \\
9000    &       1451    &       0.6     &       0.4     &       1431    &       0.4     &       0.4     &       160     &       0.8     &       0.9     &               &               &               &               &               &               &               &               &               &       845     &       0.8     &       0.8     \\
9250    &       1277    &       0.4     &       0.4     &       1300    &       0.5     &       0.4     &       159     &       2.4     &       1.1     &               &               &               &               &               &               &               &               &               &       804     &       0.6     &       0.7     \\
9500    &       490     &       1.1     &       1.3     &       428     &       1.1     &       1.4     &       156     &       1.3     &       1.1     &               &               &               &               &               &               &               &               &               &       716     &       0.5     &       0.6     \\
9750    &       361     &       1.3     &       1.5     &       347     &       1.5     &       1.6     &       136     &       3.5     &       1.4     &               &               &               &               &               &               &               &               &               &       146     &       0.2     &       0.3     \\
10000   &       1128    &       1.9     &       1.7     &       1069    &       1.7     &       1.7     &       115     &       3.6     &       1.8     &       2880    &       0.7     &       0.8     &       291     &       0.9     &       1.2     &       2403    &       1.0     &       1.1     &               &               &               \\
11000   &       631     &       2.2     &       1.7     &       641     &       2.1     &       1.9     &       115     &       7.5     &       3.1     &       2352    &       1.0     &       1.0     &       1220    &       1.3     &       1.7     &       1759    &       1.2     &       1.2     &               &               &               \\
12000   &       373     &       2.2     &       1.8     &       384     &       2.2     &       2.0     &       109     &       3.2     &       2.3     &       1719    &       1.2     &       1.3     &       1480    &       1.4     &       1.5     &       1410    &       1.5     &       1.5     &               &               &               \\
13000   &       219     &       3.0     &       2.3     &       230     &       2.8     &       2.4     &       114     &       1.9     &       2.2     &       1261    &       1.5     &       1.8     &       1051    &       1.6     &       2.0     &       1305    &       1.5     &       1.8     &               &               &               \\
14000   &       135     &       2.8     &       2.4     &       157     &       3.0     &       2.6     &       1887    &       1.0     &       1.0     &       767     &       1.7     &       2.2     &       690     &       1.8     &       2.3     &       828     &       1.7     &       2.2     &               &               &               \\
15000   &       88      &       2.3     &       2.4     &       91      &       2.5     &       3.3     &       1671    &       2.0     &       1.5     &       499     &       2.1     &       2.6     &       417     &       2.5     &       2.8     &       570     &       2.0     &       2.5     &               &               &               \\
16000   &       48      &       4.3     &       3.1     &       42      &       3.4     &       3.3     &       1330    &       1.6     &       1.7     &       374     &       2.6     &       3.0     &       327     &       2.4     &       3.2     &       388     &       2.5     &       2.9     &               &               &               \\
17000   &       29      &       4.3     &       3.5     &       37      &       2.1     &       3.6     &       761     &       1.5     &       2.1     &       304     &       2.6     &       3.4     &       235     &       3.4     &       4.1     &       305     &       2.2     &       3.1     &               &               &               \\
18000   &       17      &       1.4     &       3.3     &       17      &       6.1     &       6.2     &       413     &       2.0     &       2.5     &       218     &       3.6     &       4.1     &       190     &       3.5     &       4.6     &       223     &       3.3     &       3.8     &               &               &               \\
19000   &       4       &       8.9     &       7.3     &       3       &       13.4    &       20.0    &       307     &       2.1     &       3.0     &       202     &       3.3     &       4.0     &       157     &       4.7     &       5.5     &       185     &       3.6     &       4.5     &               &               &               \\
20000   &               &               &               &               &               &               &               &               &               &       189     &       3.2     &       4.3     &       135     &       3.4     &       5.2     &       161     &       4.1     &       4.9     &               &               &               \\
21000   &               &               &               &               &               &               &               &               &               &       189     &       4.0     &       4.8     &       142     &       5.0     &       6.6     &       153     &       3.8     &       5.3     &               &               &               \\
22000   &               &               &               &               &               &               &               &               &               &       182     &       4.9     &       4.9     &       134     &       5.1     &       6.3     &       151     &       4.7     &       5.9     &               &               &               \\
23000   &               &               &               &               &               &               &               &               &               &       138     &       3.6     &       5.0     &       146     &       6.0     &       6.5     &       139     &       5.5     &       5.9     &               &               &               \\
24000   &               &               &               &               &               &               &               &               &               &       144     &       3.3     &       4.9     &       110     &       5.3     &       6.7     &       137     &       5.3     &       6.7     &               &               &               \\
25000   &               &               &               &               &               &               &               &               &               &       135     &       4.4     &       5.9     &       106     &       4.2     &       6.0     &       98      &       4.0     &       5.6     &               &               &               \\
26000   &               &               &               &               &               &               &               &               &               &       117     &       3.9     &       6.2     &       104     &       3.7     &       5.3     &       121     &       4.3     &       5.6     &               &               &               \\
27000   &               &               &               &               &               &               &               &               &               &       100     &       3.6     &       5.4     &       105     &       5.3     &       6.3     &       109     &       5.8     &       6.6     &               &               &               \\
28000   &               &               &               &               &               &               &               &               &               &       124     &       2.8     &       4.2     &       87      &       5.9     &       8.3     &       99      &       4.9     &       7.0     &               &               &               \\
29000   &               &               &               &               &               &               &               &               &               &       102     &       2.1     &       3.2     &       69      &       4.4     &       6.6     &       85      &       6.6     &       7.6     &               &               &               \\
30000   &               &               &               &               &               &               &               &               &               &       56      &       1.1     &       1.7     &       61      &       6.7     &       8.1     &       90      &       4.4     &       6.4     &               &               &               \\
31000   &               &               &               &               &               &               &               &               &               &       12      &       0.4     &       0.6     &       88      &       6.6     &       7.7     &       110     &       4.9     &       6.2     &               &               &               \\
32000   &               &               &               &               &               &               &               &               &               &               &               &               &       73      &       6.0     &       8.2     &       96      &       4.3     &       5.5     &               &               &               \\
33000   &               &               &               &               &               &               &               &               &               &               &               &               &       57      &       6.6     &       7.7     &       84      &       5.9     &       6.1     &               &               &               \\
34000   &               &               &               &               &               &               &               &               &               &               &               &               &       59      &       6.4     &       8.1     &       78      &       4.1     &       4.4     &               &               &               \\
35000   &               &               &               &               &               &               &               &               &               &               &               &               &       38      &       8.4     &       7.3     &       53      &       3.6     &       3.8     &               &               &               \\
36000   &               &               &               &               &               &               &               &               &               &               &               &               &       34      &       6.0     &       6.9     &       31      &       4.6     &       5.2     &               &               &               \\
37000   &               &               &               &               &               &               &               &               &               &               &               &               &       35      &       5.1     &       8.6     &       30      &       5.5     &       4.8     &               &               &               \\
38000   &               &               &               &               &               &               &               &               &               &               &               &               &       22      &       6.7     &       9.2     &       15      &       3.7     &       4.1     &               &               &               \\
39000   &               &               &               &               &               &               &               &               &               &               &               &               &       13      &       5.6     &       8.5     &       17      &       13.7    &       10.0    &               &               &               \\

\hline
\multicolumn{22}{l}{The calibrations are (2) - (4) \citet{1996BaltA...5..303H} for $B_\mathrm{2}-G$;
(5) - (7) \citet{1996BaltA...5..303H} for $B_\mathrm{2}-V_\mathrm{1}$; (8) - (10) \citet{1997A&AS..122...51K};
(11) - (13) \citet{1999NewAR..43..343C} for \logg\,=\,3.5;}\\
\multicolumn{22}{l}{(14) - (16) \citet{1999NewAR..43..343C} for \logg\,=\,5.0;
(17) - (19) \citet{2005A&A...444..941P}; (20) - (22) \citet{2006A&A...458..293P}.}
\end{tabular}
\end{adjustbox}
\end{center}
\tablefoot{The first column is the \Teff\ bin. The results are shown in 
Fig. \ref{figure_binning}.
For each calibration, there is the number of stars per bin, the absolute error in [\%] and its standard
deviation in [\%].}
\end{table*}

\end{appendix}

\end{document}